\documentclass[twocol]{ametsoc} 
\usepackage{fixltx2e} 
\usepackage{ucs}              
\usepackage[utf8x]{inputenc}  

\usepackage[colorlinks,citecolor=blue,linkcolor=blue,urlcolor=blue,pdftex]{hyperref}
\usepackage{journalabbrevsAMS}

\journal{jamc} 

\newcommand{\figw}{19pc} 

\bibpunct{(}{)}{;}{a}{}{,}

\newcommand{\ROCSS}{\mathrm{ROCSS}}  
\newcommand{\BSS}{\mathrm{BSS}}      

\newcommand{\km}{\mathrm{km}  }  
\newcommand{\degr}{°}            
\newcommand{\COtwo}{CO\textsubscript{2}}

\title{Skill and reliability of seasonal forecasts for the Chinese energy sector}
\authors{Philip E. Bett\correspondingauthor{Met Office Hadley Centre, FitzRoy Road, Exeter EX1 3PB, UK.}, Hazel E. Thornton, Julia F. Lockwood}
\affiliation{Met Office Hadley Centre, FitzRoy Road, Exeter EX1 3PB, UK.}
\email{philip.bett@metoffice.gov.uk}

\extraauthor{Adam A. Scaife}
\extraaffil{Met Office Hadley Centre, FitzRoy Road, Exeter EX1 3PB, UK.\\
College of Engineering, Mathematics and Physical Sciences, University of Exeter, Exeter, UK}

\extraauthor{Nicola Golding, Chris Hewitt}
\extraaffil{Met Office Hadley Centre, FitzRoy Road, Exeter EX1 3PB, UK.}

\extraauthor{Rong Zhu, Peiqun Zhang}
\extraaffil{Laboratory for Climate Studies, National Climate Center, China Meteorological Administration, Beijing, People's Republic of China}

\extraauthor{Chaofan Li}
\extraaffil{Center for Monsoon System Research, Institute of Atmospheric Physics, Chinese Academy of Sciences, Beijing, People's Republic of China}

\abstract{We assess the skill and reliability of forecasts of winter and summer  temperature, wind speed and irradiance over China, using the GloSea5 seasonal forecast system.
Skill in such forecasts is important for the future development of seasonal climate services for the energy sector, allowing better estimates of forthcoming demand and renewable electricity supply.
We find that although overall the skill from the direct model output is patchy, some high-skill regions of interest to the energy sector can be identified.
In particular, winter mean wind speed is skilfully forecast around the coast of the South China Sea, related to skilful forecasts of the El Ni\~{n}o--Southern Oscillation. Such information could improve seasonal estimates of offshore wind power generation.
Similarly, forecasts of winter irradiance have good skill in eastern central China, with possible use for solar power estimation.
Much of China shows skill in summer temperatures, which derives from an upward trend. However, the region around Beijing retains this skill even when detrended. This temperature skill could be helpful in managing summer energy demand. 
While both the strengths and limitations of our results will need to be considered when developing seasonal climate services in the future, the outlook for such service development in China is promising.
\\
This document is \copyright{} Crown Copyright \the\year{} Met Office 
}

\begin{document}
\maketitle

\section{Introduction}
The energy sector has long been a key user of weather and climate information across timescales: short-range weather forecasts \citep[e.g.][]{Taylor2003Using, Costa2008Review}, longer-range forecasts out to several weeks ahead \citep[e.g.][]{Dubus2014Weather}, as well as projections of possible future climates decades ahead  \citep[e.g.][]{McColl2012Assessing, Wang2014Chinas}. These are all used to inform the planning, development, management and running of energy systems on those timescales.
The energy sector has also been a leader in demonstrating demand for seasonal to decadal climate prediction \citep{Buontempo2010MultiScale, BrunoSoares2015Exploring}.  In particular, seasonal forecasts of the climate in the coming 3-month period have the potential for providing real added value, in both a practical and financial sense, across a range of areas within the energy sector  \citep{Troccoli2010Seasonal, DoblasReyes2013Seasonal, Dessai2015Report, BrunoSoares2015Exploring}.  However, lack of skill in many areas of the globe has limited the uptake of this kind of information within the sector.

Energy demand is strongly related to air temperature \citep[e.g.][]{Valor2001Daily, Hor2005Analyzing, Apadula2012Relationships, Zhang2014Study, Thornton2016Role}, and the potential use of seasonal climate forecasting in demand management has been recognised for many decades \citep[e.g.][and references therein]{BrownWeiss1982Value}. The need to reduce both greenhouse gas emissions and air pollution has driven an increase in the amount of electricity supplied by renewable sources. This has, in turn, resulted in an increase in the weather-dependence of energy supply systems, and therefore in the possible utility of weather and climate forecasting to the sector.  The energy sector in China faces similar issues as other countries, although particular features are the increases in demand due to rapid urbanisation \citep{Wang2014Effects, Lin2014Energy}, and a recent large growth in both installed and planned renewable energy capacity \citep[e.g.][]{Hong2013Assessment, CNREC2014China, Lo2014Critical, Qiang2016Strategic}.

Seasonal forecasts can give an early warning of a season of high demand -- such as a particularly cold winter or hot summer --  or of reduced supply, due to low wind speeds, more cloudy/hazy periods reducing solar power generation, or low water levels (which can affect both hydroelectric plants and the cooling systems for traditional thermal power plants).  Predicting conditions that could damage energy infrastructure, such as storms, could also be valuable.   In all these cases, seasonal forecasts can enable mitigation plans to be put in place: storing more water in dams, rescheduling maintenance work, making early decisions around staff availability and financial planning for the coming 3-month period.  The information could be used by a range of people, including  industry regulators, network operators, energy production companies, maintenance contractors, and financial market traders.

However, seasonal forecasts are most useful if they have sufficient skill to allow decision-making.  Furthermore, what `sufficient' means (beyond being statistically significant) will depend on the particular use case.  If  forecasts are not sufficiently skillful, then while an organisation might be happy to receive forecast information, they might not be able to make a decision based on it.

While skillful forecasts for some variables in some parts of the world have been possible for some time \citep[e.g.][]{Arribas2011GloSea4}, recent advances in seasonal forecasting systems have led to major improvements in the skill of extratropical features such as the North Atlantic Oscillation (NAO, e.g. \citealt{Athanasiadis2014Representation, Athanasiadis2016Multisystem, Butler2016Climatesystem, Smith2016Seasonal} and references therein). \cite{Scaife2014Skillful} demonstrated skill in NAO forecasts from version 5 of the Met Office's Global Seasonal forecasting system, GloSea5 \citep{MacLachlan2015Global}. This  has led to the development of seasonal climate services for various sectors in the UK, including hydrology \citep{Svensson2015Longrange}, transport \citep{Palin2016Skilful} and energy \citep{Clark2017Skilful}.  While seasonal forecasting has a long history in China, the traditional low skill from dynamical models has led to a wide literature in statistical downscaling and statistical forecasting \citep[e.g.][]{Wang2015Review, Xing2016Longlead}.  It is timely therefore to examine the skill of the GloSea5 system in China, for direct forecasts of energy-relevant climate variables. The results  could allow the development of future climate services \citep{Golding2017ImprovingDraft}, that is, the provision and use of climate information to enable better informed decisions.

The development and use of such climate services has become a major undertaking world-wide, with international coordination being facilitated by the Global Framework for Climate Services \citep[GFCS,][]{Hewitt2012Global}. The GFCS focuses on 5 priority sectors, one of which is the energy sector.  China is developing its own framework aligned to the GFCS, called the China Framework for Climate Services. This  brings together actors involved in scientific research, climate service development, service providers and users, to ensure that available capability and services meet  users' needs.

In this paper we assess the skill of seasonal forecasts of wind speed, irradiance, and temperature across China, from the GloSea5 system, and consider the implications  for the wind power, solar power and energy demand sectors.  We firstly describe the data sets and methods used in section~\ref{s:datamethods}. We then present our results in section~\ref{s:results}, considering a China-wide overview of each variable before focusing on some specific areas of interest.  We discuss our conclusions in section~\ref{s:concs}.


\section{Data and analysis methods}\label{s:datamethods}

\subsection{Data sets}\label{s:datasets}
In this paper we use the hindcast data set produced to assess the version of GloSea5 that was deployed operationally at the Met Office in February 2015.  This is based on the second \emph{Global Coupled} configuration (GC2) of the HadGEM3 global climate model, described in  \cite{Williams2015The}.  HadGEM3-GC2 uses the GA6.0 configuration of the Met Office Unified Model (UM, version 8.4) as its atmospheric component, on an N216 grid\footnote{i.e. $432$ cells east--west by $324$ cells north--south.} (a horizontal resolution of $0.83\degr$ in longitude and $0.55\degr$ in latitude) and 85 vertical levels reaching a height of $85\,\km$ near the mesopause \citep{Walters2016Met}. This is  coupled to the GL6.0 configuration of the JULES land surface model \citep{Best2011Joint}, the GO5.0 configuration of the NEMO ocean model with a  $0.25\degr$ nominal resolution and 75 vertical levels (version 3.4, \citealt{Megann2014GO50, Madec2008nemo}), and the GSI6.0 configuration of the CICE sea ice model (version 4.1, \citealt{Rae2015Development, Hunke2010cice}).  The GloSea5 system is described in full by \cite{MacLachlan2015Global} and references therein.

The assessment hindcast was produced to examine the skill of the system in forecasting for winter (December--January--February, DJF) and summer (June--July--August, JJA) only.  Lagged ensemble `forecasts'  were initialised on three start dates centred on 1st November (for DJF) and 1st May (for JJA) for each of the 20 years of the hindcast data set, producing a total of 24 members for each hindcast season.  The DJF hindcasts cover (boreal) winter 1992/1993 to winter 2011/2012, and the JJA hindcasts cover (boreal) summers 1992--2011.  The details of the initialisation are described in \cite{MacLachlan2015Global}.

Our ability to robustly assess  forecast skill is limited by the size of the hindcast, both in terms of the number of years and the number of members.  The operational GloSea5 forecast system uses 42 members, rather than the 24 available in the hindcast used here, so the probability distributions inferred using the hindcast will be less well resolved than they would be operationally. Furthermore, it has been shown that the skill itself depends directly on the size of the ensemble, as it allows better identification of predictable signals  \citep{Scaife2014Skillful, Eade2014Do, Dunstone2016Skilful}.  For our purposes, the ensemble size means that we can regard levels of skill shown here, where significant, to be lower limits of the actual skill that could be realized in the operational system.  The robustness of the skill estimates is also limited by the number of years in the hindcast \citep{Kumar2009Finite}, as this places a restriction on the number of different types of event that are sampled in the period of study.  The impact of the limited  hindcast period is quantified by assessing the statistical significance of the correlations between the hindcast and observations, described in the next subsection.

We use the ERA-Interim reanalysis data \citep{Dee2011ERAInterim} as a proxy for observations in this paper.  While ERA-Interim and other reanalyses are frequently used to validate temperature and wind data from climate models, their use as a proxy for irradiance observations is more contested \citep[e.g.][]{Boilley2015Comparison}.  We have compared some of our irradiance results against the SARAH-E satellite-derived observational data \citep{datasetSARAHErad, Amillo2014New}, and find that, for the seasonal means averaged over large areas that we consider here, and in the standardised units we use, ERA-Interim compares very well. However, the use of climatological aerosols in both ERA-Interim and SARAH-E, and indeed GloSea5, means that the impact of aerosols on interannual variability  remains an important uncertainty.

\subsection{Skill assessment methodology}\label{s:skillmethods}
In this paper we focus on three meteorological variables of interest to the energy sector in China: near-surface air temperature (related to energy demand); 10-metre wind speed (linked to wind power generation); and downwelling shortwave irradiance at the surface (related to solar power generation).  Precipitation is also of great importance for the energy sector, as China has a very large, and growing, hydroelectric industry.  \cite{Li2016Skillful} have already shown that GloSea5 has significant skill in forecasting summer precipitation in the Yangtze river basin, where the Three Gorges Dam is located, and this has led to the development of a trial forecast service \citep{Golding2017ImprovingDraft}.  Further work by \cite{Lu2017SkillfulDraft} found high levels of skill in GloSea5 forecasts of winter precipitation, over south-east China.

We take the following approach to skill assessments.  For each variable, we first map the Pearson correlation between the hindcast ensemble mean and the observations.  While the limited timespan and ensemble size of the hindcast means that forecasts from single grid cells (or even small regions) are likely to be very noisy and not robust, these maps give a good indicative overview, and provide context when selecting larger geographical areas of interest for subsequent analysis. These regions are selected based on their interest to the energy sector: we consider  the current and likely future development of substantial energy supply or demand, and if they appear to have some promising skill (regions are not selected on the basis of skill maps alone)

We then assess the skill in each region in more detail. We consider 3 types of plot, each with an associated skill score:
\begin{enumerate}
\item Standardised\footnote{By standardised we mean that we subtract the average then divide by the standard deviation $\sigma$. The result is an anomaly time series in units of $\sigma$.} time series (using the hindcast ensemble mean); with the Pearson correlation $r$.
\item Reliability and sharpness diagrams, with the Brier skill score $\BSS$. These show the joint distribution of hindcast probabilities and observed frequencies for a particular class of event, and the $\BSS$ measures how much better the forecast system is compared to using climatology in that case (see Appendix~A for further details).
%
\item ROC diagrams, with the ROC skill score $\ROCSS$. These describe  the ability of the forecast system to distinguish between events occurring or not occurring (see Appendix~B for further details).
\end{enumerate}
In all cases, we follow the standard \cite{WMO2010} procedure for assessing such forecasts.  In particular, this means weighting by the cosines of the grid cell latitudes when aggregating grid cells in the region in question.  

A Student's t-test is often used to assess whether a Pearson sample correlation is significantly non-zero.  This assumes that the variables in question are independent in time and have a Gaussian distribution.  While this will only be approximately true in our case, we nevertheless use the t-test to give an \emph{indicative} measure of significant skill in our correlation maps, to aid (but not determine) selection of regions of interest: we draw a contour around areas that would be significantly non-zero at the $5\%$ level if the assumptions of the test applied exactly.\footnote{In our case of having $20$ years of data, the threshold in correlation corresponding to the $5\%$ significance level is $|r| > 0.44$.}  We are not correcting for multiple testing here, so it should be expected that some of the regions marked as notionally significant will be false positives; again, the significance contour should not be taken as definitive.

There are some cases where there is a clear trend running through the data.  The reproduction of such a trend by the hindcast is a genuinely useful aspect of skill, as it shows that the forecast system is capable of maintaining the impact of whatever forcing caused the trend, after being initialised.  However, it can hide information about the ability of the model to evolve correctly away from its initialised state more generally, which is also of interest when assessing the model.  We have therefore also looked at the correlation skill after detrending, which we perform by simply removing the linear least-squares regression fits to the hindcast ensemble mean and observational time series, separately.  Note that we are not making any assumptions as to the cause or significance of any trends.  The time series is sufficiently short that natural interannual and decadal-scale variability will be very important, even before considering anthropogenic drivers of climate change such as \COtwo\ emissions, land-use change (affecting effective surface roughness and hence wind speed) and aerosol emissions\footnote{Although ERA-Interim uses climatological aerosols, it assimilates other observations that could be forced by anomalous aerosol emissions, making it particularly complicated to assess in this way.} (affecting surface irradiance and temperature).   We simply remove the empirical linear trend.

Our reliability and ROC diagrams are made in terms of probabilistic forecasts of particular types of event: we consider the probability of the variable in question being above the median, in the upper/middle/lower tercile, or in the top/bottom quintile, of its historical distribution. These quantiles are calculated for the hindcast and observational data sets independently, from their own climatologies. This means that our reliability diagrams and Brier skill scores are insensitive to a simple bias in the mean state between the two data sets.

For each type of event (e.g. upper tercile), the distribution of ensemble members each year\footnote{We use cross-validation when calculating the quantile values, with a window length of 1 year (i.e. 1 DJF or JJA period), following  \cite{WMO2010}: The quantile in question is calculated separately for each year, from the 19 years of data remaining after that year is masked out.} provides the forecast probability of that event occurring, in each grid cell.  Using bins of probability with width 0.1, we then consider all the years when the event was forecast to occur with probability in a given bin, and count the frequency of times when it was observed to actually occur. The counting is done in each grid cell, and we then pool the counts from all the grid cells in the region using cos-latitude weighted sums, to calculate the skill scores and reliability/ROC plots.

\section{Results}\label{s:results}
Here we show maps of the correlation between GloSea5 and ERA-Interim for each variable.  In each case, we then go on to examine particular regions in more detail, through their regional time series, reliability and ROC diagrams.

\subsection{Wind speed}
\begin{figure}
\includegraphics[width=\figw]{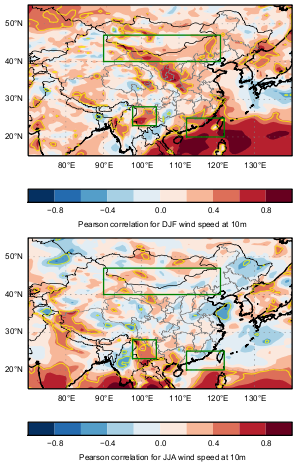} 
   \caption{Maps of the correlation of seasonal mean 10m wind speed between GloSea5-GC2 hindcast and ERAI, for winter (top) and summer (bottom). Green boxes highlight regions discussed in the text, and yellow contours mark correlations $|r| > 0.44$, suggestive of significant skill. }
  \label{f:windcormaps}
\end{figure}

Maps of the correlation between ERA-Interim and the GloSea5 hindcast for 10\,m wind speeds are shown in Figure~\ref{f:windcormaps}, for winter and summer. Maps of the correlation of detrended time series are practically indistinguishable (not shown).

\begin{figure}
\includegraphics[width=\figw]{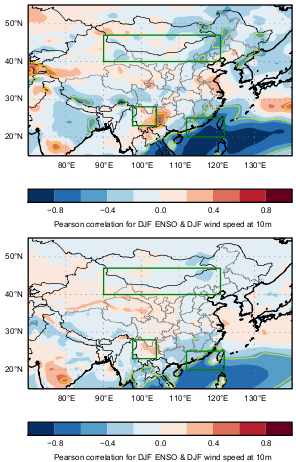} 
\caption{Maps of the correlation of the DJF Nino3.4 index and 10\,m wind speeds, in ERA-Interim (top) and in the ensemble members of the GloSea5-GC2 hindcast (bottom). Note that we do not use the ensemble mean here, as we are showing the internal correlation within the model: correlating the model's ensemble means with each other would result in a stronger signal, as it would  remove natural noise that is still present in the single `realisation' of ERA-Interim.}
\label{f:ensocormaps}
\end{figure}

While there are some areas of significant positive skill for wind speed in DJF, they are rather patchy.  A major highlight however is the very high skill in the South China Sea, off the south and south-east coasts of China, with some skill being retained inland. This is likely to be  related to the skill in forecasting the El Ni\~{n}o--Southern Oscillation ($r\approx0.9$ for the DJF Ni\~{n}o3.4 index,\footnote{The Ni\~{n}o3.4 index is the timeseries of sea surface temperature anomalies in the region 120°W--170°W, 5°S--5°N.} see \citealt{MacLachlan2015Global}), and in its teleconnections over China: these are shown in Figure~\ref{f:ensocormaps} in terms of  correlations between the Ni\~{n}o3.4 index and wind speed.  While the overall response in the region differs in detail between GloSea5 and ERA-Interim,  the significant anticorrelation in the South China Sea is present in both, with weaker wind speeds correlated to El Ni\~{n}o events.  In this region, as part of the East Asian winter monsoon, there is strong north-easterly flow around the south eastern edge of the Siberian--Mongolian High \citep{Chang2006Asian}. During El Ni\~{n}o events, there is increased subsidence over the Maritime Continent, increasing the surface pressure over that region and thus reducing the land--sea pressure gradient and resulting monsoonal winds \citep{Zhang1996Impact}.

\begin{figure*}\centering
\includegraphics[width=\textwidth]{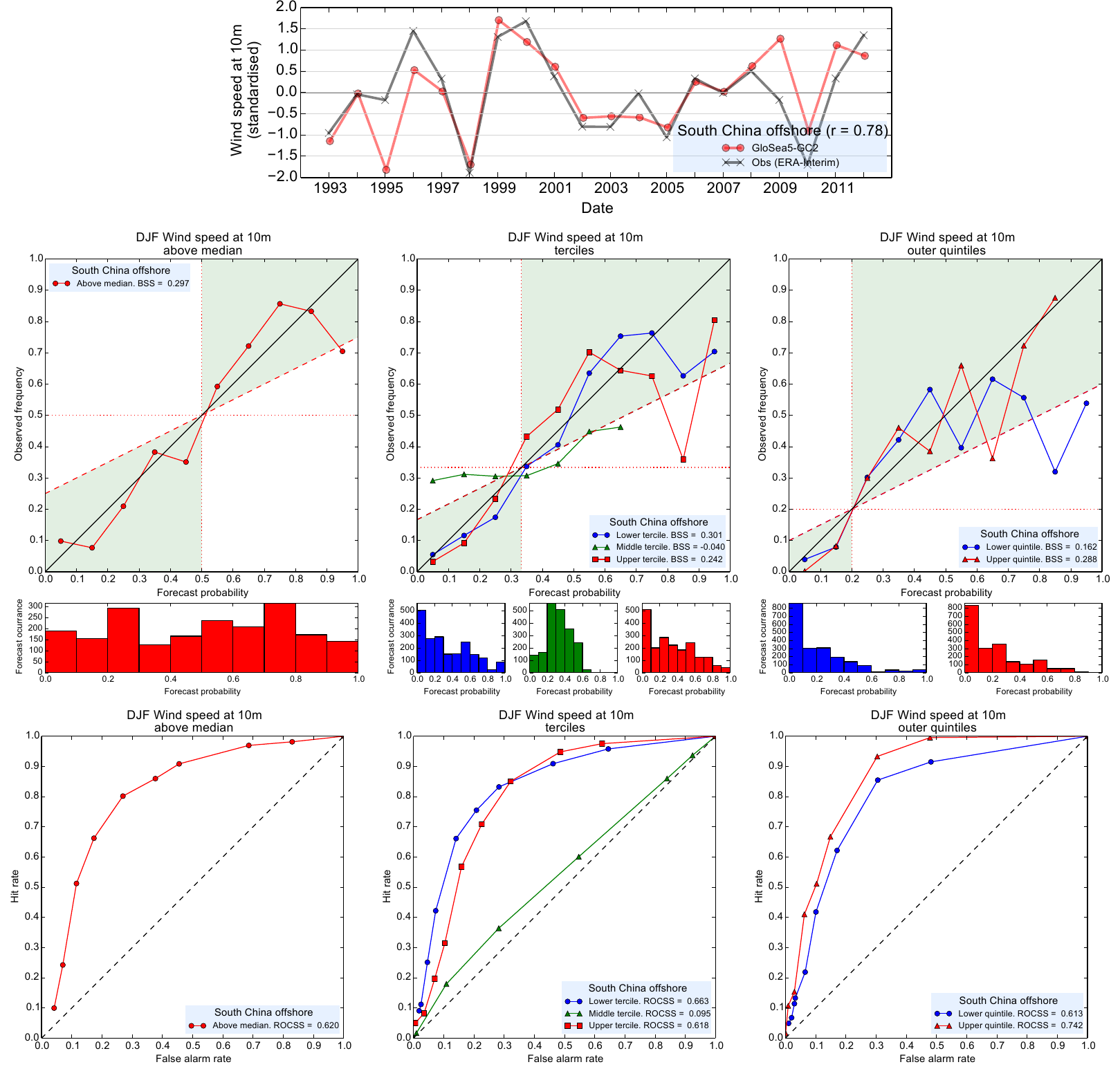} 
\caption{Skill and reliability assessment plots for winter wind speeds in the South China Sea region marked in Figure~\protect\ref{f:windcormaps}.  Top: standardised time series, using the ensemble mean of the seasonal hindcast.  The other rows show the reliability \& sharpness diagrams (middle; see Appendix~A for details) and ROC plots (bottom, see Appendix~B), for the three event definitions: above median winds (left), terciles (centre), and outer quintiles (right), as labelled.  In each panel, the corresponding skill score is given in the legend.  }
\label{f:DJFwindSOffshoretwocol}
\end{figure*}

Figure~\ref{f:DJFwindSOffshoretwocol} shows skill and reliability for winter wind speeds in this South China Sea region (the south-eastern green box in Figure~\ref{f:windcormaps}). While the deterministic ensemble-mean forecast has a correlation of $r\approx 0.8$, the reliability diagrams show that this region also exhibits skillful, reliable and sharp probabilistic forecasts of above-median wind speed events.  Results are also very good for upper and lower tercile events, and upper and lower quintiles, although these are more noisy.  This is reflected in the sharpness diagrams: high-probability forecasts of outer quintile events in particular are very poorly sampled by the hindcast, whereas forecasts of above-median events are well sampled across the full range of probabilities.  It is also important to note that this skill in wind speeds is \emph{not} retained in summer (as seen in the lower panel of Figure~\ref{f:windcormaps}): in that case, the skill scores $r=0.13$, $\BSS=-0.07$ and $\ROCSS=0.004$ are not significantly different to zero (not shown).

\begin{figure}
\includegraphics[width=\figw]{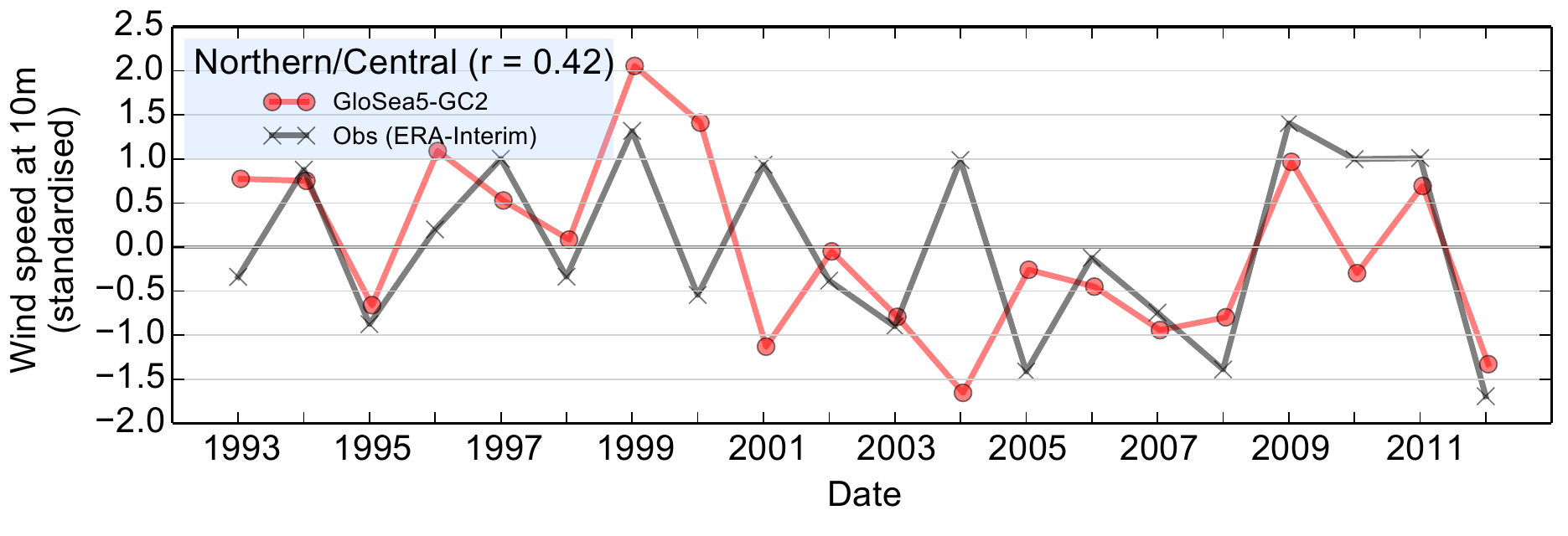} 
\caption{Standardised time series for winter wind speeds in the northern-central China/southern Mongolia region (the northern box marked in Figure~\protect\ref{f:windcormaps}). The correlation is marked in the legend.} 
\label{f:DJFwindNCentral}
\end{figure}

Some other regions also appear to have reasonably high levels of skill.  Figure~\ref{f:DJFwindNCentral} shows the timeseries for winter wind speed in north-central China and southern Mongolia (northern green box in Figure~\ref{f:windcormaps}).  This is a region of particularly high wind resource \citep[e.g.][]{CNREC2014China,Davidson2016Modelling}, so being able to forecast it could be of great practical use.  However, the skill in this particular region is marginal ($r=0.42$).  Research is ongoing to see if statistical models, based on larger-scale atmospheric drivers such as the Arctic Oscillation and Middle Eastern Jet Stream \citep{Yang2004Upstream, He2017Impact}, could result in higher levels of skill than the direct model output.

\begin{figure}
\includegraphics[width=\figw]{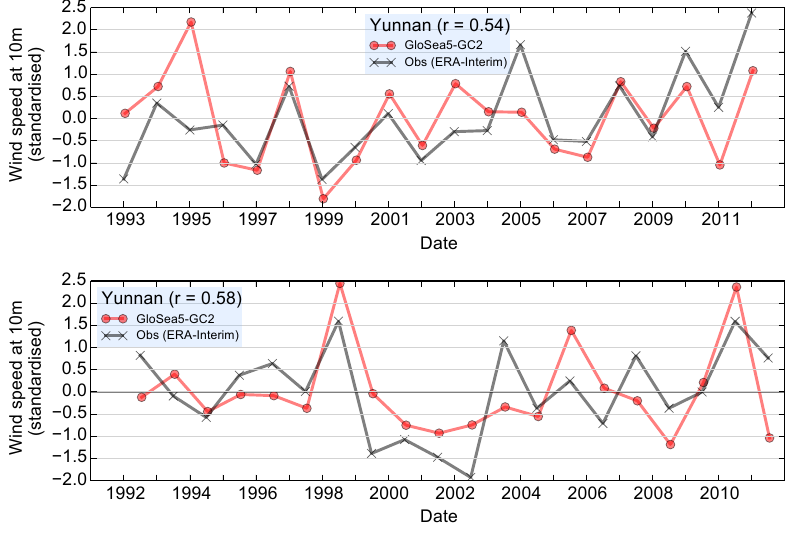} 
\caption{Standardised time series for winter (top) and summer (bottom) wind speeds in the Yunnan region (the south-west box marked in Figure~\protect\ref{f:windcormaps}). The correlation is marked in the legend.} 
\label{f:DJFJJAwindYunnan}
\end{figure}

Yunnan province,  in southern China, shows modest but significant skill for wind speeds, in both winter and summer (Figure~\ref{f:DJFJJAwindYunnan}; the region is also marked in Figure~\ref{f:windcormaps}).  Yunnan is very mountainous, and energy production has traditionally been dominated by hydroelectricity. However, in recent years there has been substantial drive to utilise the available wind resource \citep{Liang2015Research}, and a seasonal forecast could prove useful.


\subsection{Irradiance}\label{s:irradiance}
\begin{figure}
\includegraphics[width=\figw]{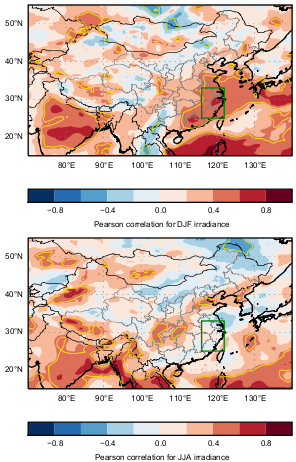} 
   \caption{Maps of the correlation of seasonal mean irradiance between GloSea5-GC2 hindcast and ERAI, for winter (top) and summer (bottom).  As in Figure~\protect\ref{f:windcormaps}, contours mark regions of notionally significant skill.  The green box highlights the eastern/coastal region discussed in the main text. }
  \label{f:irradcormaps}
\end{figure}

Figure~\ref{f:irradcormaps} shows correlation maps for irradiance in winter and summer.  In winter, there is a broad area of promising skill in eastern China and the East China Sea.  This bears a strong resemblance to the patterns of skill in winter precipitation shown by \cite{Lu2017SkillfulDraft}, perhaps unsurprisingly as both rainfall and irradiance are strongly related to cloudiness. \cite{Lu2017SkillfulDraft} determined that the key drivers of precipitation predictability here are ENSO, and rainfall in the eastern Indian Ocean/Bay of Bengal.  It is reasonable to assume that the same processes that affect winter precipitation in this region also affect cloudiness and therefore downwelling shortwave irradiance at the surface.

\begin{figure*}\centering
\includegraphics[width=\textwidth]{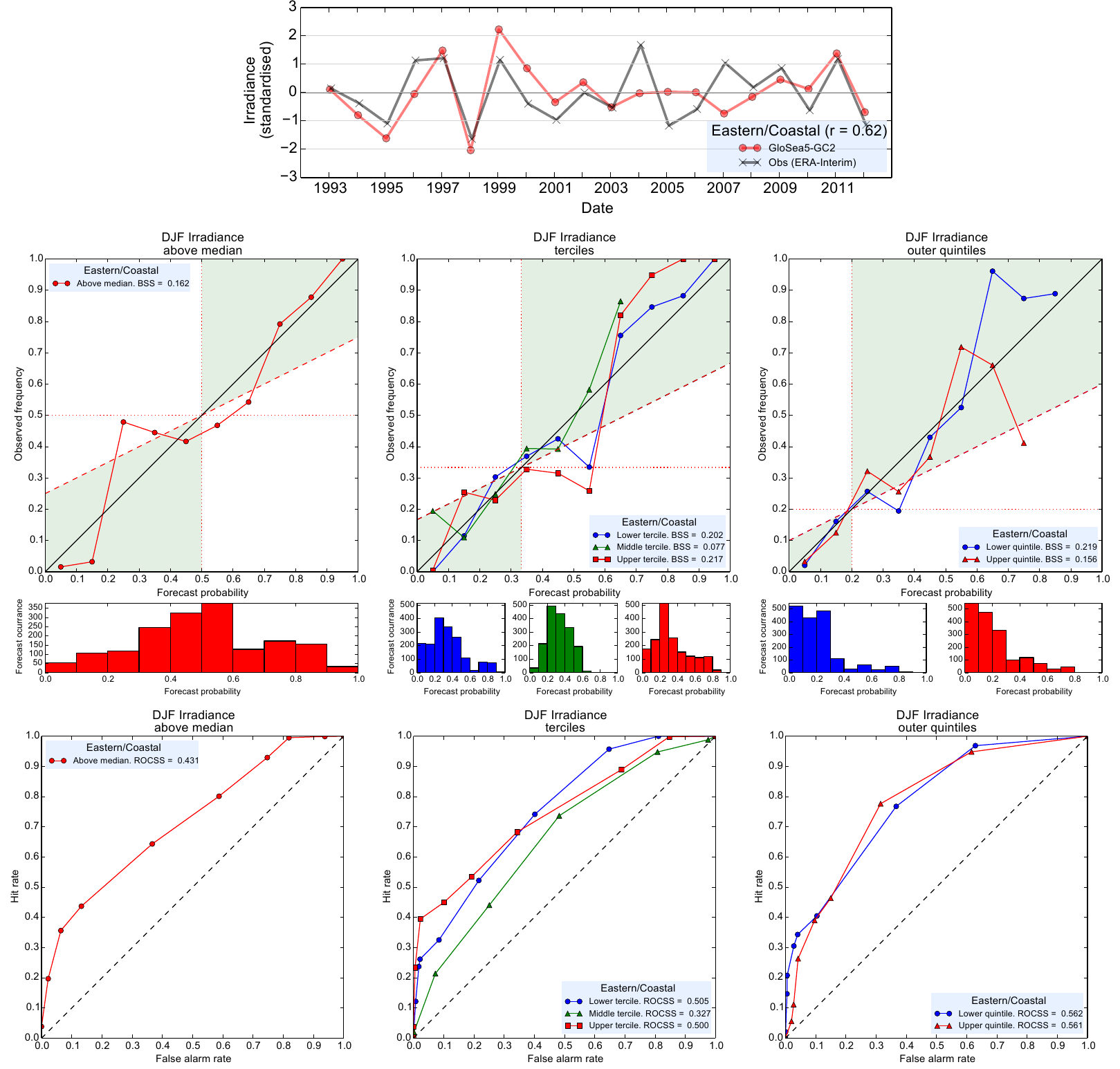} 
\caption{Skill and reliability assessment of winter irradiance, for the easternmost coastal region of China marked with a green box in Figure~\protect\ref{f:irradcormaps}. As in Figure~\protect\ref{f:DJFwindSOffshoretwocol}, we show the standardised time series using the ensemble mean (top), with reliability and sharpness diagrams (middle) and ROC plots (bottom),  for above-median (left), tercile (centre) and outer quintile (right) events. Each panel shows its corresponding skill score, and further details of the reliability and ROC plots are given in Appendix~A and~B respectively.} 
\label{f:DJFirradECoastal}
\end{figure*}

Figure~\ref{f:DJFirradECoastal} shows the winter irradiance skill in the eastern China region in more detail, using the eastern green box marked in Figure~\ref{f:irradcormaps}: probabilistic forecasts remain reasonably skillful and reliable for tercile and outer quintile events. While this isn't a region of particularly high solar radiation resource within China \citep{CNREC2014China}, it is an area of high population density with many urban centres, including Shanghai.  The potential for high levels of demand modulated by large numbers of roof-mounted solar panels means that being able to forecast winters with more or less solar generation than usual could be of value.

In summer, the correlation overall across China is much poorer, with the  eastern China region considered above now having $r=0.03$, consistent with zero (not shown). Regions further west appear to have higher levels of skill, but it is still vary patchy.

As already discussed, since both ERA-Interim and GloSea5 use climatological aerosols, we are unable to assess the impact of interannual aerosol variability on seasonal irradiance forecasts. This might have a strong impact in urban areas affected by haze for example, and it remains an important uncertainty when considering the application of these results to solar power generation.


\subsection{Temperature}
\begin{figure*}\centering
\includegraphics[width=\textwidth]{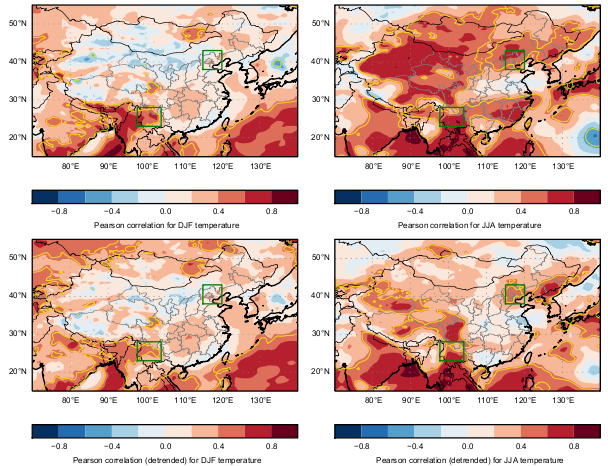} 
\caption{Maps of the correlation of seasonal mean air temperature between GloSea5-GC2 hindcast and ERAI, for winter (left) and summer (right).  The bottom panels show the results for detrended data.  As in Figure~\protect\ref{f:windcormaps}, contours mark regions of notionally significant skill. The northern (Beijing area) and southern (Yunnan province) green boxes are discussed in the text. }
  \label{f:tempcormaps}
\end{figure*}

Figure~\ref{f:tempcormaps} shows correlation maps for temperature, including a comparison with detrended data.  It is clear that there is significant skill in predicting summer  temperatures over large areas of China, and that in many regions these are due to positive trends over the hindcast period.  The winter temperatures are less affected by trends, and indeed show very little skill overall.  As with winter wind speed, research is underway to improve the forecasts through the use of larger-scale atmospheric drivers.

One exception is Yunnan province in south--central China, which we also  highlighted for wind speed skill: here it shows positive skill for both winter and summer temperatures.  Yunnan is less urban than many more eastern parts of China, so the utility of a temperature forecast in energy demand planning is more limited.  However, agriculture and tourism are both very important for Yunnan, which could benefit from a skillful  seasonal temperature forecast.

\begin{figure*}\centering
\includegraphics[width=\textwidth]{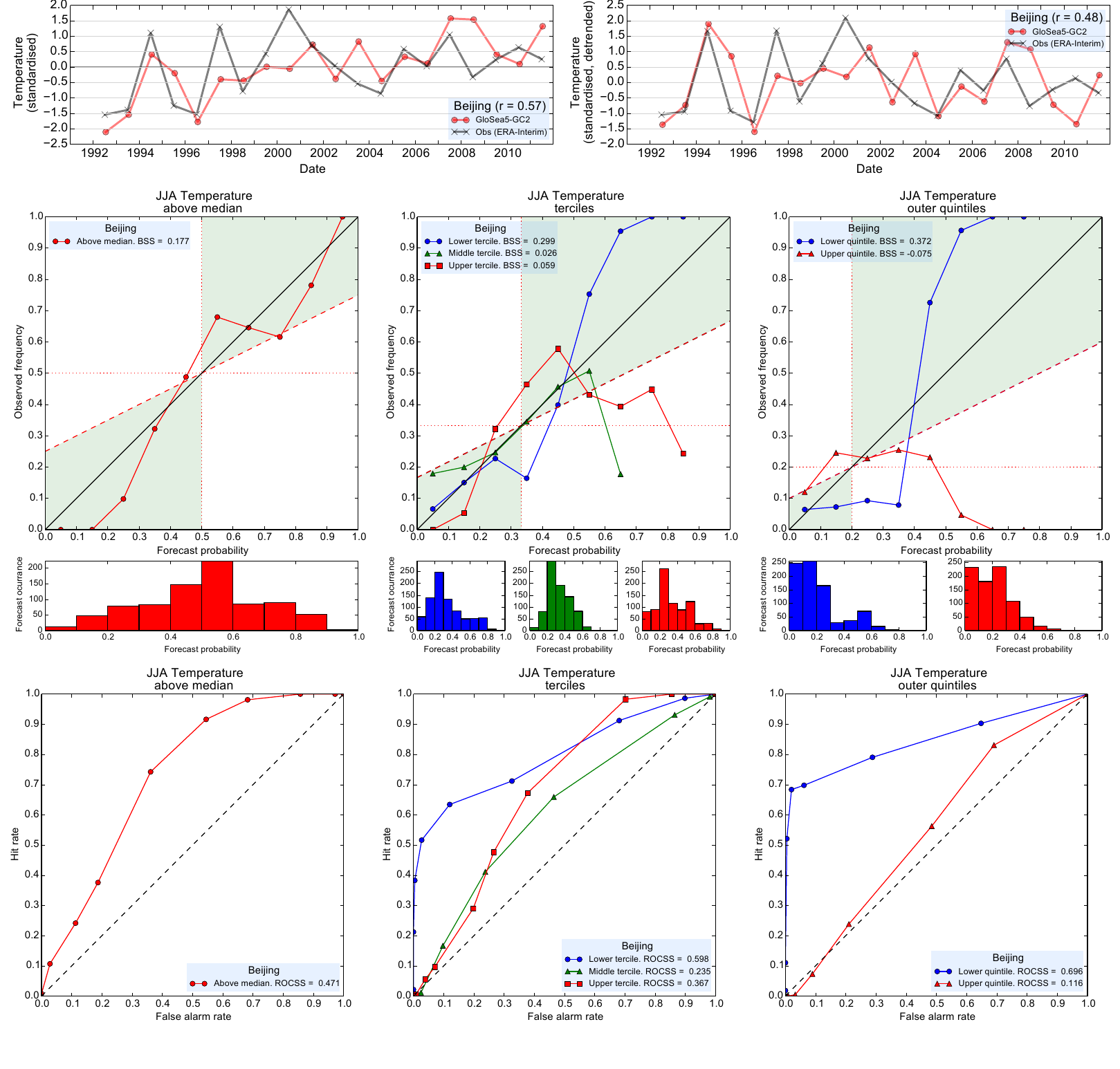} 
\caption{Skill and reliability assessment for summer temperatures in Beijing (as marked with a box in Figure~\protect\ref{f:tempcormaps}), following Figures~\protect\ref{f:DJFwindSOffshoretwocol} and~\protect\ref{f:DJFirradECoastal}.  We include the timeseries of both the standard (top-left) and detrended (top-right) data; the other results are produced without detrending.} 
\label{f:JJAtempBeijing}
\end{figure*}

Clearly, the potential for useful seasonal forecasts of energy demand, and hence temperature, is greatest in urban centres. In particular, energy demand in Beijing is strongly related to temperature in summer \citep{Zhang2014Study}. It is important therefore that the region around Beijing (the northern green box in Figure~\ref{f:tempcormaps}) also shows some skill for summer temperatures, before and after detrending, with correlations of $\sim0.5$--$0.6$ (Figure~\ref{f:JJAtempBeijing}).    There is less skill however for probabilistic forecasts more detailed than the simple `above-average' case, although this might be improved by looking at a larger region, reducing statistical noise.


\section{Discussion and conclusions}\label{s:concs}
Our results have shown that, while overall skill for energy-relevant variables in China remains patchy, some specific areas have significant skill: winter wind speeds in the South China Sea, winter solar irradiance in eastern/southern China, and summer temperatures across much of China (due to the trend), including Beijing (even when detrended).  Taken together with similarly-promising results for skillful summer precipitation in the Yangtze river basin \citep{Li2016Skillful}, and winter precipitation in southeastern China \citep{Lu2017SkillfulDraft}, there are clear opportunities to develop useful seasonal climate services for specific cases within China.

Indeed, taking our results and those on precipitation skill together, there are clear potential climate service applications beyond the energy sector: for example, forecasting risks to  agriculture and transport, and risks of flooding.


We have only considered the skill of the direct model output from the GloSea5 hindcast here.  This represents a minimum level of forecast skill, in two senses. Firstly, the operational forecast ensemble is larger than that available in this hindcast, and it is well-established that the forecast skill in GloSea5 increases with the size of the ensemble \citep[e.g.][]{Scaife2014Skillful, Li2016Skillful, Dunstone2016Skilful}.  

Secondly, statistical models linking larger-scale drivers directly to the impact variable of interest may offer further improvement in predictability \citep[e.g][]{Scaife2016North}.  This technique has been used for seasonal forecasts in the UK  \citep[e.g.][]{Svensson2015Longrange, Palin2016Skilful, Clark2017Skilful}, and is often used already in China \citep[e.g.][]{Xiao2012Progress, Wang2013Subtropical, Peng2014Seasonal, Wang2015Review, Xing2016Longlead}.  Research is  ongoing to understand the predictability of larger-scale drivers in GloSea5, and how they can be used to improve sector-specific forecasts.  

Furthermore, the most user-relevant services are likely to be forecasts of the particular impact of interest to the user, such as energy supply or demand.  A next step in developing seasonal climate services based on these results should therefore be to assess the skill of GloSea5 against such direct impacts data, where available from a potential user.   The way that forecasts are communicated and handled also affects the usefulness of the forecast \citep[e.g.][]{Taylor2015Report,Davis2016Barriers}: user engagement is therefore key to optimising a climate prediction service.

Nevertheless, if co-developed with users and communicated carefully, our results show some areas of very promising skill, allowing the development of improved, skillful seasonal climate forecasting services for specific parts of the energy sector, and other sectors, in China.

\acknowledgments
This work and its contributors (PB, JL, AS, CH and NG) were supported by the UK--China Research \& Innovation Partnership Fund through the Met Office Climate Science for Service Partnership (CSSP) China as part of the Newton Fund.  
HT was supported by the Joint UK BEIS/Defra Met Office Hadley Centre Climate Programme (GA01101).
CL was supported by the National Natural Science Foundation of China (Grant No. 41320104007).
 PB would like to thank  Hongli Ren, Jo Camp,  Robin Clark and Margaret Gordon for helpful discussions.


\appendix[A]
\appendixtitle{Reliability and sharpness diagrams, and Brier skill scores $\BSS$}\label{a:reliab}
Model reliability is a description of how closely the forecast probabilities of an event correspond to the frequency of that event being observed in historical data, assessing a conditional bias in the forecast system: for example, we might find that every time the event is forecast to occur with $70\%$ probability, it actually occurs only $60\%$ of the time.  Having characterised such discrepancies, they can then be removed through calibration, resulting in improved forecasts.  A full description can be found in \cite{Wilks2011Forecast}, but we describe the key points for interpreting our plots here.

The reliability diagram for a given class of event is a plot of the observed frequency of the event, at times when  it was forecast to occur with a given probability. As described in section~\ref{s:datamethods}\ref{s:skillmethods}, we use bins of probability of width 0.1, and pool the event counts from all grid cells in the chosen region of interest.  For the set of years where the event was forecast to occur with a given probability, we plot the on the vertical axis the fraction \emph{of those years} when the event actually occurred, and join the points from each bin with a line.

We mark additional lines in our reliability diagrams (sometimes called an attributes diagram, \citealt{Hsu1986Attributes}).  The 1:1 line  (black, solid) marks ``perfect reliability'', differentiating between underconfident and overconfident forecasts -- these will have steeper or shallower reliability lines respectively than the `perfect' case.  The climatological frequency (e.g. $1/3$ for terciles) is marked with a dotted horizontal ``no resolution'' line: If the reliability line lies along this, then it cannot resolve different events into different probabilities, as all forecasts occur at the climatological rate.  A line midway between ``perfect reliability'' and the ``no resolution'' line is called the ``no skill'' line (dashed), as only points above this line make a positive contribution to the Brier skill score.  We shade this region of skill in green.

The Brier skill score measures how much better the forecast system is relative to climatology,\footnote{In general, the Brier skill score compares the forecast system to any reference forecast, but here we use climatology as the reference.} and can be written as
\newcommand{\REL}{\mathrm{REL}}
\newcommand{\RES}{\mathrm{RES}}
\newcommand{\UNC}{\mathrm{UNC}}
\begin{equation}
\BSS = \frac{\RES - \REL}{\UNC}.
\end{equation}
Here, the \emph{resolution} $\RES$ is the weighted mean square distance between the points and the ``no resolution'' line, the \emph{reliability} $\REL$ is the weighted mean square distance between the points and the ``perfect reliability'' line, and the ``uncertainty'' $\UNC$ is the product of the observed climatological frequency and its complement, e.g. $\frac{1}{3}\times\frac{2}{3}$.  The skill is positive if $\RES>\REL$, i.e. if the points in the reliability line are closer to ``perfect reliability'' than to the ``no resolution'' line.

Below each reliability diagram, we include a sharpness diagram, a histogram of the distribution of forecasts made in each probability bin.  If the histogram is flat, then the hindcast has sampled the full range of possible forecast probabilities and is described as \emph{sharp}.  If it is strongly peaked at the climatological frequency for the event, then the system has no sharpness and mostly just predicts climatology.  Taken together, the sharpness and reliability diagrams provide a complete description of the joint distribution of observed frequencies and forecast probabilities.


\appendix[B]
\appendixtitle{ROC diagrams and ROC skill scores $\ROCSS$}\label{a:roc}
\newcommand{\HR}{\mathrm{HR}}
\newcommand{\FAR}{\mathrm{FAR}}
\newcommand{\Nhits}{N_\mathrm{H}}
\newcommand{\Nmisses}{N_\mathrm{M}}
\newcommand{\Nfalse}{N_\mathrm{FA}}
\newcommand{\Ncorrej}{N_\mathrm{CR}}
Relative Operating Characteristic (ROC) diagrams describe how well the forecast system can distinguish between classes of event occurring and not occurring \citep[again, see][for a more full description]{Wilks2011Forecast}. In practice, we construct ROC diagrams and scores using the same event classes and probability bins as used for the reliability diagrams, counting events and performing weighted sums over contributing grid cells. Four aggregates of the event counts are made, for each probability bin $p$:
\begin{itemize}
\item $\Nhits(p)$, the number of hits: the number of times the event was forecast with probability $> p$, and observed to occur.
\item $\Nmisses(p)$, the number of misses: the number of times the event was observed, but was not forecast with probability $> p$.
\item $\Nfalse(p)$, the number of false alarms: the number of times the event was forecast with probability $> p$, but was not observed to occur.
\item $\Ncorrej(p)$, the number of correct rejections: the number of times the event was not observed to occur, and was not forecast with probability $> p$.
\end{itemize}
We then calculate:
\begin{itemize}
\item Hit Rate, $\HR(p) = \Nhits / (\Nhits + \Nmisses) $
\item False Alarm Rate, $\FAR(p) = \Nfalse / (\Nfalse + \Ncorrej)$
\end{itemize}
The ROC diagram then is a plot of Hit Rate against the False Alarm Rate, for a series of probability thresholds. A skillful system has $\HR>\FAR$ and therefore a ROC curve in the top-left of the diagram; the 1:1 line (dashed in our plots)  delineates no skill, as $\HR=\FAR$.  We therefore use the area under the ROC curve $A_\mathrm{ROC}$ as a measure of skill, and scale it to produce a skill score that lies between 0 (no skill) and 1 (perfect):
\begin{equation}
\ROCSS = 2\times A_\mathrm{ROC} - 1.
\end{equation}
ROC diagrams are insensitive to calibration of the forecast probabilities, so complement the reliability diagrams -- they assess the potential usefulness of the forecast system after calibration.

\bibliographystyle{ametsoc2014}
\bibliography{cssp_energyskill}

\end{document}